\documentclass[utf8]{frontiersFPHY}
\usepackage{physics}
\usepackage{url,hyperref,lineno,microtype}
\usepackage{subcaption}
\usepackage[onehalfspacing]{setspace}

\def\orcid#1{\kern .08em\href{https://orcid.org/#1}{\includegraphics[keepaspectratio,width=0.7em]{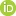}}}

\setcitestyle{square} 

\def\firstAuthorLast{L. Coraggio, S. Pastore, and C. Barbieri} 
\def\Authors{Luigi Coraggio\,$^{1}$, Saori Pastore\,$^{2}$, Carlo Barbieri\,$^{3,4,5}$\orcid{0000-0001-8658-6927}}
\def\Address{
$^{1}$Istituto Nazionale di Fisica Nucleare, Sezione di Napoli, 80126 Napoli, Italy\\
$^{2}$ Physics Department and McDonnell Center for the Space
Sciences at Washington University in St. Louis, St. Louis, Missouri
63130, USA\\
$^{3}$Department of Physics, University of Surrey, Guildford GU2 7XH, United Kingdom\\
$^{4}$Dipartimento di Fisica, Universit\`a degli Studi di Milano, Via Celoria 16, I-20133 Milano, Italy\\
$^{5}$INFN, Sezione di Milano, Via Celoria 16, I-20133 Milano, Italy}

\def\corrAuthor{~}
\def\corrEmail{~}

\begin{document}
\onecolumn
\firstpage{1}

\title{Editorial: The  Future of Nuclear Structure: Challenges and Opportunities in the
  Microscopic Description of Nuclei}


\author[\firstAuthorLast ]{\Authors} 
\address{} 
\correspondance{} 
\extraAuth{}

\maketitle

 
The past two decades have witnessed tremendous progress in the
microscopic description of atomic nuclei. Within this approach, nuclei 
are described in terms of nucleons interacting via realistic two- and 
three-body forces, constrained to accurately reproduce a 
large body of data for few nucleons systems. The goal of the nuclear 
theory community is to gain an accurate and predictable understanding 
of how the properties of many-body systems, along with their dynamics 
and structure, emerge from internucleon correlations induced by the 
strong interaction. 

Progress in the microscopic (or, \emph{ab initio}) theory has been quite notable
and it has been supported by two major pillars: First, thanks to the advent of
Effective Field Theories (EFTs), we can now systematically develop nuclear
Hamiltonians that are rooted in the fundamental properties and symmetries of 
the underlying theory of QCD. Second, advances in computational resources and
novel powerful algorithms allow us to solve \emph{i}) the many-nucleon problem 
efficiently, and \emph{ii}) quantify the degree of reliability of theoretical 
calculations and predictions. In many cases, microscopic computations achieve
an accuracy that is comparable or superior to the precision delivered by 
current EFT interactions. This sparked a renewed interest to further broaden 
the focus of \emph{ab initio} theory and address open problems in nuclear physics.

While the status of the first pillar has been recently discussed by
``The Long-Lasting Quest for Nuclear Interactions: The Past, the Present 
and the Future" Topical Review on this Journal, here we focus on the exciting
new developments in microscopic theory.
At present, \emph{ab initio} computations of nuclear structure include up to 
medium-mass isotopes. The heaviest systems currently reached---with different 
degrees of accuracy---have mass number $A \approx 140$. These computational
limits are constantly being pushed forward. At the same time, the community is 
expanding into new directions, in particular towards the study of  electroweak 
observables and nuclear reactions, that nowadays require predictions with an 
accuracy never reached before for similar mass ranges.

In collecting the contributions for this Research Topic, we sought to gather contributions
from authors who could summarize the current state-of-the-art microscopic calculations 
in Nuclear Theory, favouring a selected but broad view over an attempt to cover every 
application. All presented contributions stem from well-established methods in computational
nuclear structure, and indicate recent theoretical advances and prospective outlooks, challenges
and opportunities for Nuclear Theory.  Most importantly, it is our hope that this collection
will confer a `big picture’, including references to basic material, that will be valuable
for young researches who intend to enter this exciting discipline.

The richness of applications in modern \emph{ab initio} nuclear theory can be appreciated in
\href{https://doi.org/10.3389/fphy.2020.00379}{\emph{\color{teal} Hergert}}'s contribution that provides us with a general overview of the 
most successful microscopic many-body approaches currently in use~\cite{10.3389/fphy.2020.00379}. Traditionally, the refinement 
and sophistication of these computational tools has given fundamental support to advance the
theories of nuclear forces. Quantum Monte Carlo (QMC) techniques allow to solve the many-body
Schr\"odinger equation with high accuracy for light nuclei up to masses A$\sim$16-40. 
\href{https://doi.org/10.3389/fphy.2020.00117}{\emph{\color{teal} Gandolfi et al.}} discuss the use of QMC methods 
(namely, Variational, Green's Function and Auxiliary Diffusion Monte Carlo methods) in combination
with local chiral interactions in coordinate space~\cite{10.3389/fphy.2020.00117}. QMC methods are used in lattice effective 
field theory, where the EFT Lagrangian is implemented in momentum space with nucleons and pions 
placed on a lattice. \href{https://doi.org/10.3389/fphy.2020.00174}{\emph{\color{teal} Lee}} discusses the basic features of this approach 
and its high potential for understanding clustering phenomena~\cite{10.3389/fphy.2020.00174}.

For heavier isotopes, \emph{ab initio} theories can be pushed to masses A$\sim$140 
provided that one retains only the relevant nuclear excitations, as it is done 
through all-orders resummations. Among these methods, the self-consistent Green's function
(SCGF) theory gives direct access to the spectral information probed by a wide range of 
experiments as reviewed in detail by \href{https://doi.org/10.3389/fphy.2020.00340}{\emph{\color{teal} Som\`a}}'s contribution~\cite{10.3389/fphy.2020.00340}. 
Once in the region of the nuclear chart that corresponds to medium masses, open shell 
isotopes become the next challenge to be addressed by the theory. In fact, resolving 
the degeneracy in uncorrelated systems requires large scale configuration mixing. 
\href{https://doi.org/10.3389/fphy.2020.00345}{\emph{\color{teal} Coraggio and Itaco}} demonstrate how this can be handled by projecting 
the correlated many-body states into a shell model Hamiltonian, using the so-called ``Q-box"
formalism~\cite{10.3389/fphy.2020.00345}. A similar strategy is shared by other computational frameworks, such as coupled 
cluster and in-medium SRG, that are touched upon in the contribution by \href{https://doi.org/10.3389/fphy.2020.00379}{\emph{\color{teal} Hergert}}~\cite{10.3389/fphy.2020.00379}.
A less conventional approach to open shells is to  break SU(1) symmetry (in short, allowing 
for breaking particle number conservation). This is discussed by \href{https://doi.org/10.3389/fphy.2020.00340}{\emph{\color{teal} Som\`a}} 
within SCGFs~\cite{10.3389/fphy.2020.00340} and by \href{https://doi.org/10.3389/fphy.2020.00164}{\emph{\color{teal} Tichai et al.}} in the framework of many-body perturbation theory~\cite{10.3389/fphy.2020.00164}.

The remainder of this topical review focuses on selected open challenges in Nuclear Theory
that require an \emph{ab initio} approach. Two contributions show different aspect of studying
infinite nucleon systems and the implications for astrophysical scenarios. \href{https://doi.org/10.3389/fphy.2020.00153}{\emph{\color{teal} Tews}} 
covers QMC calculations of the equation of state (EoS) of dense matter in neutron stars~\cite{10.3389/fphy.2020.00153}. With
the recent observation of star mergers and the birth of multi-messenger astronomy, it has become 
of prime importance to understand the finite temperature properties of the EoS. 
\href{https://doi.org/10.3389/fphy.2020.00387}{\emph{\color{teal} Rios}} discusses this topic and how the structure of neutron matter
depends on temperature, using SCGF theory~\cite{10.3389/fphy.2020.00387}.

In the quest for physics beyond the Standard Model, Nuclear Theory, and in particular accurate calculations
of neutrino-nucleus interactions at all energy scaler, plays a crucial role. This is carefully analyzed by 
\href{https://doi.org/10.3389/fphy.2020.00116}{\emph{\color{teal} Rocco}}'s contribution  that address this challenge with emphasis on impacts to 
neutrino oscillations experimental programs~\cite{10.3389/fphy.2020.00116}.
The last contribution of this Topical Review addresses one of the hardest open challenges in the interpretation of
experimental data: the lack of a truly first-principles theory that can describe \emph{consistently} both structure 
and reaction processes. \href{https://doi.org/10.3389/fphy.2020.00285}{\emph{\color{teal} Rotureau}} highlights recent steps in deriving an \emph{ab inito}
optical potential using the coupled cluster method (that, together with SCGF, is one of the two possible 
approaches to this problem)~\cite{10.3389/fphy.2020.00285}.

We are really grateful to all the scientists participating in this project and hope that the reader will enjoy this Topical Review.  

\bibliographystyle{frontiersinHLTH&FPHY}
\bibliography{biblio}

\begin{thebibliography}{10}
\expandafter\ifx\csname natexlab\endcsname\relax\def\natexlab#1{#1}\fi
\expandafter\ifx\csname urlstyle\endcsname\relax
  \expandafter\ifx\csname doi\endcsname\relax
  \def\doi#1{doi:\discretionary{}{}{}#1}\fi \else
  \expandafter\ifx\csname doi\endcsname\relax
  \def\doi{doi:\discretionary{}{}{}\begingroup \urlstyle{rm}\Url}\fi \fi
\expandafter\ifx\csname selectlanguage\endcsname\relax
  \def\selectlanguage#1{}\fi

\bibitem[{Hergert(2020)}]{10.3389/fphy.2020.00379}
Hergert H.
\newblock A guided tour of ab initio nuclear many-body theory.
\newblock {\em Frontiers in Physics\/} {\bf 8} (2020) 379.
\newblock \doi{10.3389/fphy.2020.00379}.

\bibitem[{Gandolfi et~al.(2020)Gandolfi, Lonardoni, Lovato, and
  Piarulli}]{10.3389/fphy.2020.00117}
Gandolfi S, Lonardoni D, Lovato A, Piarulli M.
\newblock Atomic nuclei from quantum monte carlo calculations with chiral eft
  interactions.
\newblock {\em Frontiers in Physics\/} {\bf 8} (2020) 117.
\newblock \doi{10.3389/fphy.2020.00117}.

\bibitem[{Lee(2020)}]{10.3389/fphy.2020.00174}
Lee D.
\newblock Recent progress in nuclear lattice simulations.
\newblock {\em Frontiers in Physics\/} {\bf 8} (2020) 174.
\newblock \doi{10.3389/fphy.2020.00174}.

\bibitem[{Somà(2020)}]{10.3389/fphy.2020.00340}
Somà V.
\newblock Self-consistent green's function theory for atomic nuclei.
\newblock {\em Frontiers in Physics\/} {\bf 8} (2020) 340.
\newblock \doi{10.3389/fphy.2020.00340}.

\bibitem[{Coraggio and Itaco(2020)}]{10.3389/fphy.2020.00345}
Coraggio L, Itaco N.
\newblock Perturbative approach to effective shell-model hamiltonians and
  operators.
\newblock {\em Frontiers in Physics\/} {\bf 8} (2020) 345.
\newblock \doi{10.3389/fphy.2020.00345}.

\bibitem[{Tichai et~al.(2020)Tichai, Roth, and
  Duguet}]{10.3389/fphy.2020.00164}
Tichai A, Roth R, Duguet T.
\newblock Many-body perturbation theories for finite nuclei.
\newblock {\em Frontiers in Physics\/} {\bf 8} (2020) 164.
\newblock \doi{10.3389/fphy.2020.00164}.

\bibitem[{Tews(2020)}]{10.3389/fphy.2020.00153}
Tews I.
\newblock Quantum monte carlo methods for astrophysical applications.
\newblock {\em Frontiers in Physics\/} {\bf 8} (2020) 153.
\newblock \doi{10.3389/fphy.2020.00153}.

\bibitem[{Rios(2020)}]{10.3389/fphy.2020.00387}
Rios A.
\newblock Green's function techniques for infinite nuclear systems.
\newblock {\em Frontiers in Physics\/} {\bf 8} (2020) 387.
\newblock \doi{10.3389/fphy.2020.00387}.

\bibitem[{Rocco(2020)}]{10.3389/fphy.2020.00116}
Rocco N.
\newblock Ab initio calculations of lepton-nucleus scattering.
\newblock {\em Frontiers in Physics\/} {\bf 8} (2020) 116.
\newblock \doi{10.3389/fphy.2020.00116}.

\bibitem[{Rotureau(2020)}]{10.3389/fphy.2020.00285}
Rotureau J.
\newblock Coupled-cluster computations of optical potential for medium-mass
  nuclei.
\newblock {\em Frontiers in Physics\/} {\bf 8} (2020) 285.
\newblock \doi{10.3389/fphy.2020.00285}.

\end{thebibliography}

\end{document}